# The Realistic Interpretation of Quantum Mechanics

Roumen Tsekov

Department of Physical Chemistry, University of Sofia, 1164 Sofia, Bulgaria

It is demonstrated how quantum mechanics is generated by stochastic momentum kicks from the force carriers, transmitting the fundamental interactions between the point particles. The picture is consistent with quantum field theory and points out that the force carriers are the only quantum particles. Since the latter are waves in the coordinate space, they are responsible for the wavy character of quantum mechanics.

The Schrödinger equation[1] is fundamental for physics and chemistry. It Is proposed to reflect the controversial de Broglie idea that quantum particles can behave as waves as well, resembling the Einstein photons. The wave-particle duality is unable, however, to answer many relevant physical questions. For example, if an electron is a wave distributed in space, its negative charge must be divided somehow into many small pieces but such particles with infinitesimal fractional charges are never observed. Moreover, if the electron localizes back as a point particle, the related work against the electrostatic repulsion between the fractional charges would tend to infinity. Keeping the Schrödinger equation as experimentally proven, we are going to explore another interpretation of quantum mechanics, where the particles remain points at any time as in classical mechanics. It will be demonstrated that quantum mechanics is merely due to the force carriers transmitting the fundamental interactions. While in a previous paper[2] we modeled them as point particles, now the force carriers are waves/quasiparticles in the coordinate/momentum spaces, respectively. This is the reason for the wavy character of quantum mechanics, not the point particles themselves.

## A Hydrogen Atom

For the sake of simplicity let us consider first a hydrogen atom, constructed from an electron with mass $m_1$ and a proton with mass $m_2$, but the conclusions will be generalized later. The relevant Schrödinger equation describes the evolution in time of the wave function $\psi(r_1, r_2, t)$ at any positions $r_1$ and $r_2$ of the electron and proton, respectively,

$$i\hbar\partial_t\psi = -\hbar^2\partial^2_{r_1}\psi/2m_1 - \hbar^2\partial^2_{r_2}\psi/2m_2 + u_{12}\psi \qquad (1)$$

For the further analysis it is crucial to present the Coulomb interaction potential in the Fourier form as

$$u_{12} \equiv -\frac{e^2}{4\pi\varepsilon_0|r_1-r_2|} = -\int_{-\infty}^{\infty}\frac{e^2}{\varepsilon_0 k^2}\exp[ik\cdot(r_1-r_2)]d^3(k/2\pi) \qquad (2)$$

Because there is no wave function in classical mechanics, it is appropriate to employ the Wigner function[3] $W(p_1,r_1,p_2,r_2,t)$, being the quasi-probability density in the particles phase space. Via the Wigner-Weyl quantization[4] one can calculate all statistical properties in quantum mechanics. The Wigner function is a useful tool and its marginal probability densities coincide with the exact solutions of the Schrödinger equation both in coordinate and momentum representations. The evolution of the Wigner function follows directly from the Schrödinger equation (1) and obeys the Wigner-Liouville equation[5,6]

$$\partial_t W + p_1\cdot\partial_{r_1}W/m_1 + p_2\cdot\partial_{r_2}W/m_2 = 2u_{12}\sin[\hbar(\overleftarrow{\partial}_{r_1}\cdot\vec{\partial}_{p_1} + \overleftarrow{\partial}_{r_2}\cdot\vec{\partial}_{p_2})/2]W/\hbar \qquad (3)$$

where the arrows show the direction of the operator action. For the sake of the further analysis, it is important to replace here the interaction potential via its Fourier form (2) to obtain

$$\partial_t W + p_1\cdot\partial_{r_1}W/m_1 + p_2\cdot\partial_{r_2}W/m_2 = \int_{-\infty}^{\infty}\frac{2e^2\exp[ik\cdot(r_1-r_2)]}{i\hbar\varepsilon_0 k^2}\sinh[\hbar k\cdot(\vec{\partial}_{p_1}-\vec{\partial}_{p_2})/2]Wd^3(k/2\pi) \qquad (4)$$

The hyperbolic sine function is a difference of two exponential terms, which represent translation operators of momenta in Eq. (4). Hence, one can rewrite it in a decisive form

$$\partial_t W + p_1\cdot\partial_{r_1}W/m_1 + p_2\cdot\partial_{r_2}W/m_2 = \int_{-\infty}^{\infty}\frac{e^2\exp[ik\cdot(r_1-r_2)]}{\varepsilon_0 ik\cdot\hbar k}$$
$$[W(p_1+\hbar k/2,r_1,p_2-\hbar k/2,r_2,t) - W(p_1-\hbar k/2,r_1,p_2+\hbar k/2,r_2,t)]d^3(k/2\pi) \qquad (5)$$

Suddenly, the physics behind the Schrödinger equation becomes transparent. The large expression in the brackets describes transfer of momentum $\hbar k$ from the proton to the electron by a single photon. The other term in the collision integral (5) is the acting force per unit transmitted momentum, i.e. it is the characteristic frequency of photon exchange. The ratio between the collision and photon frequencies $2\alpha/\pi$ is proportional to the fine structure constant. The integration over positive and negative wave vectors $k$ accounts for momenta transfers in both directions. The photon momentum $\hbar k$ is the only quantum quantity in Eq. (5). The point charges themselves are not quantum and do not propagate as waves; they simply swim in the quantum sea of the photons. This realistic picture correlates well to quantum field theory, stating that the fundamental forces are transmitted by force carriers.[7] One can easily recognize a virtual photon Feynman propagator in the term $-1/k^2$ of Eq. (2). The latter seems not quantum, since the photon mass is zero, and it is stationary, because the retardation is neglected. Obviously, a photon is wave in the coordinate space but it behaves as a quasiparticle in the momentum space, as discovered by Einstein. Therefore, the force carriers are the reason for the wavy character of quantum mechanics. Because the Feynman propagator governs the force carriers of all fundamental interactions, Eq. (2) can be easily modified to describe gravity and strong forces as well. This is the reason why all fundamental interaction potentials vanish inversely proportional to the distance between the point particles.[7] In general, the force carriers transmit either energy or momentum, which create the interaction potential and generate the quantum motion. The force carriers are the de Broglie pilot waves and the Bohm hidden variables.

If the momenta of the point particles are much larger than the force carriers' quanta, one can expand in a $\hbar k$-series the expression in the brackets in Eq. (5). Keeping the leading terms only yields the classical Liouville equation

$$\partial_t W + p_1 \cdot \partial_{r_1} W / m_1 + p_2 \cdot \partial_{r_2} W / m_2 = \partial_{r_1} u_{12} \cdot \partial_{p_1} W + \partial_{r_2} u_{12} \cdot \partial_{p_2} W \tag{6}$$

Therefore, the importance of quantum effects depends on the momenta of the point particles, not on their masses only. Of course, heavy particles possess large momenta even at slow velocity and that is why they obey often classical mechanics. It is evident that a free electron should obey Eq. (6) without an external potential, since there are no force carriers without interaction. Hence, the free electron trajectory $x = x_0 + p_0 t / m$ increases linearly in time. However, to fix the electron at the initial position, one must apply a strong initial attractive potential at $x_0$. According to Eq. (5) this initial potential will generate momenta kicks $\pm \hbar k / 2$ randomizing the electron momentum. Once the initial potential is switched off at $t = 0$, the electron will propagate freely with an initial momentum acquired by the photon kicks. Since $<p_0> = 0$ and the initial variables $<x_0 p_0> = 0$ are uncorrelated, the evolution of the free electron position dispersion $\sigma_x^2 = \sigma_x^2(0) + \sigma_p^2(0)t^2/m^2$ will grow quadratic in time. Due to the Fourier transform, the initial momentum fluctuation $\sigma_p(0) = \hbar\sigma_k(0)/2 = \hbar/2\sigma_x(0)$ is related to the width $\sigma_x(0)$ of the initial potential. Thus, the Wigner function for a free electron is a Gaussian probability density, which is spreading via the well-known law $\sigma_x^2 = \sigma_x^2(0) + (\hbar t/2m)^2/\sigma_x^2(0)$, and the Heisenberg inequality holds at any time due to of the Fourier transformation properties.

## A System of Many Point Particles

Because the fundamental interactions are pairwise additive, the total potential energy is a sum $U = \sum u_{ij}$ of pair potentials like (2). Thus, the many-particles problem reduces straightforward to the same physical picture, emerging from discrete force carriers. Let us consider now a general system of N particles with 3N-dimensional vectors of positions $r$ and momenta $p$, respectively. The Schrödinger equations in the coordinate and momentum representations read

$$i\hbar\partial_t\Psi = -\hbar^2\partial_r \cdot M^{-1} \cdot \partial_r\Psi/2 + U(r)\Psi \qquad i\hbar\partial_t\Phi = p \cdot M^{-1} \cdot p\Phi/2 + U(i\hbar\partial_p)\Phi \qquad (7)$$

where $M$ is the 3Nx3N diagonal mass matrix. The Fourier image of the interaction potential defines the total propagator of the force carriers

$$\tilde{U}(k) = \int_{-\infty}^{\infty} U(r)\exp(-ik \cdot r)d^{3N}r \qquad (8)$$

and represents the interaction energy density distributed over the force carriers' momenta as well. As was shown, quantum mechanics is a result of collisions in the momentum space. Hence, the relevant momentum Schrödinger equation (7), accomplished by Eq. (8), acquires naturally an integral form

$$i\hbar\partial_t\Phi = p \cdot M^{-1} \cdot p\Phi/2 + \int_{-\infty}^{\infty} \tilde{U}(k)\Phi(p - \hbar k, t)d^{3N}(k/2\pi) \qquad (9)$$

which seems more appropriate for complex calculations than the differential one. Moreover, if the point particles are relativistic, one can easily replace the kinetic energy $p \cdot M^{-1} \cdot p / 2$ via the corresponding expression from the special relativity to derive a relativistic Schrödinger equation.

The Wigner function is a partially inverted Fourier image of the momentum density matrix[3]

$$W(p,r,t) \equiv \int_{-\infty}^{\infty} \Phi(p - \hbar k / 2, t) \bar{\Phi}(p + \hbar k / 2, t) \exp(ik \cdot r) d^{3N}(k/2\pi) \tag{10}$$

Its evolution obeys the Wigner-Liouville equation,[5,6] which is traditionally presented by an operator of the potential energy,

$$\partial_t W + p \cdot M^{-1} \cdot \partial_r W = 2U \sin[\hbar(\bar{\partial}_r \cdot \vec{\partial}_p)/2]W/\hbar = [U(r + i\hbar \partial_p / 2) - U(r - i\hbar \partial_p / 2)]W / i\hbar \tag{11}$$

One can rewrite Eq. (11) in an alternative form, however, symbolizing the collisions from Eq. (5),

$$\partial_t W + p \cdot M^{-1} \cdot \partial_r W = [W(p + i\hbar \partial_r / 2, r, t) - W(p - i\hbar \partial_r / 2, r, t)]U / i\hbar \tag{12}$$

Expressing now the potential $U$ by the corresponding Fourier integral (8) yields after some rearrangements a classically looking alternative

$$\partial_t W + p \cdot M^{-1} \cdot \partial_r W = \int_{-\infty}^{\infty} \tilde{U} \partial_r \exp(ik \cdot r) \cdot \partial_p W_k d^{3N}(k/2\pi) \qquad W_k \equiv \int_{-1/2}^{1/2} W(p + \xi \hbar k, r, t) d\xi \tag{13}$$

As expected, all quantum effects are due to the force carriers' momenta $\hbar k$ randomizing the momenta of the point particles.[8] The normalized probability density $W_k$ offers another interpretation as a random phase approximation of momenta kicks of the force carriers on the point particles. The Bayesian product $\tilde{U}(k)W_k(p,r,t)$ represents the joint distribution of the interaction energy over the force carriers' momenta in the point particles phase space. Again, when $p \gg \hbar k/2$ then $W_k \approx W$ and Eq. (13) reduces logically to the classical Liouville equation

$$\partial_t W + p \cdot M^{-1} \cdot \partial_r W = \partial_r U \cdot \partial_p W \tag{14}$$

The accuracy of this asymptote depends strongly on the interaction energy Fourier distribution $\tilde{U}$. The unexpected validity of Eq. (14) for harmonic oscillators is due to the fact that the force carriers are also harmonic vibrations. Hence, there is no mode-mode coupling due to linearity of the underlying dynamics. As in the case of a free particle, the quantum effects of a harmonic oscillator originate solely from the initial distribution. For instance, the displacement of a harmonic oscillator with an own frequency $\omega_0$ reads $x = x_0 \cos(\omega_0 t) + p_0 \sin(\omega_0 t)/m\omega_0$ and, if the initial displacement and momentum are uncorrelated, the position dispersion is $\sigma_x^2 = \sigma_x^2(0)\cos^2(\omega_0 t) + \sigma_p^2(0)\sin^2(\omega_0 t)/m^2\omega_0^2$. The initial momentum fluctuation $\sigma_p(0) = \hbar \sigma_k(0)/2 = \hbar/2\sigma_x(0)$ can be expressed from the initial potential. The corresponding position dispersion $\sigma_x^2 = \sigma_x^2(0)\cos^2(\omega_0 t) + (\hbar/2m\omega_0)^2 \sin^2(\omega_0 t)/\sigma_x^2(0)$ reduces to the well-known expression $\sigma_x^2 = \hbar/2m\omega_0$, if one is looking for a stationary solution. The oscillator energy is injected by the initial potential during the initial positioning. In contrast to a free particle, $\hbar\omega_0/2$ is universal due to stationarity of the harmonic oscillator. Because of the parametric resonance, the oscillator can absorb external photons with frequency $\omega_0$ only and the possible oscillator energies are $(n+1/2)\hbar\omega_0$. Hence, there is no way to emit the zero-point energy $\hbar\omega_0/2$.

An extensive discussion goes in the literature about the Wigner function positivity.[9-11] In general, $W$ must possess negative values at some points due to the orthogonality of the stationary solutions of the Schrödinger equation.[12] For the hydrogen atom $W \geq 0$ is positive in the ground state,[13,14] as well as for a harmonic oscillator. Our expectation is that the Wigner function must be always positively defined in the ground state, while in the excited states $W$ can be negative somewhere. The reason for this is that the excited states appear only at the presence of external photons. Hence, the latter will exercise additional momenta kicks on the electron, which are not included in Eq. (5) so far. As was demonstrated, the Schrödinger equation is generated by the exchange of force carriers between the interacting particles. Since there are few gauge bosons, Eq. (13) is appropriate only for systems with fundamental interactions and the total potential energy must be a superposition of Feynman's momenta propagators. If one employs artificial or even approximate potentials, they would correspond to unphysical propagators of non-existing force carriers, which could result in inconsistent solutions of the Schrödinger equation. It seems that a negative value of the Wigner function is a smart indicator for problems with potentials and initial or boundary conditions. For example, the so-called cat state of a single particle possesses a bimodal Wigner function.[11] Its positivity problem is due, however, to unphysical initial conditions, since it is impossible to fix at the beginning a single point particle at two different places at once.

## The Underlying Stochastic Dynamics

It is interesting how the point particles are moving after all. Borrowing ideas from classical electrodynamics, the Lagrangian of the entire system can be written in the form

$$L = \dot{R} \cdot M \cdot \dot{R}/2 - U(R) + A(R) \cdot \dot{R} \tag{15}$$

where $R(t)$ and $\dot{R}(t)$ are the 3N-dimensional vectors of the real trajectories and velocities of the point particles, respectively. Due to energy conservation the Lagrangian (15) does not depend explicitly on time. Apart from the scalar potential $U$, the 3N-dimensional vector potential $A$ accounts for the N force carriers. This is consistent with quantum field theory,[7] where the vector potential is the wave function of virtual particles in the Klein-Gordon equation. Because the traditional magnetic forces are neglected in Eq. (7), due to their relativistic character, $A$ is independent of the point particles' velocities $\dot{R}$. Thus, the point particles momenta $P \equiv \partial_{\dot{R}} L$ consist of a differentiable part $M \cdot \dot{R}$ and momenta kicks $A$,[15] while the corresponding Euler-Lagrange equation reads

$$\dot{P} = \partial_R L = -\partial_R U + \partial_R A \cdot \dot{R} \tag{16}$$

One can recognize the fluctuation force pumping momenta via kicks in the last term. Introducing the particles momenta $P = M \cdot \dot{R} + A$ in Eq. (16) yields the corresponding stochastic Newton equation

$$M \cdot \ddot{R} + \dot{R} \cdot \partial_R A + \partial_R U = \partial_R A \cdot \dot{R} \tag{17}$$

which resembles the Brownian motion.[15] In contrast to standard dissipation, however, the friction tensor of the dissipative force $\dot{A}$ is stochastic and $\partial_R A$ possesses zero mean value to avoid any entropy production. The stochasticity in Eq. (17) originates from the fluctuations of $A$, being a random function of the configuration $R$, i.e. the evolution of the point particles positions causes instant redistribution of the force carriers. The 3N-dimensional Lorentz force $\partial_R A \cdot \dot{R} - \dot{R} \cdot \partial_R A$ does no work, since it is always orthogonal to the velocity $\dot{R}$. Thus, the energy $E \equiv P \cdot \dot{R} - L = \dot{R} \cdot M \cdot \dot{R}/2 + U$ remains constant, explicitly independent of $A$. To respect the Ehrenfest theorem, the mean value of the stochastic Lorentz force must be zero.

The phase space probability density $W = <\delta(r-R)\delta(p-P)>$ is averaged over the stochastic realizations of the vector potential and it is positively defined everywhere. Differentiating $W$ on time and substituting the Euler-Lagrange equation (16) yields

$$\partial_t W + p \cdot M^{-1} \cdot \partial_r W = \partial_p \cdot <\delta(r-R)\delta(p-P)(\partial_R U - \partial_R A \cdot \dot{R})> \qquad (18)$$

In the derivation of this equation the term $\partial_r \cdot M^{-1} \cdot <\delta(r-R)\delta(p-P)A> = 0$ is set zero, because the force carriers' waves are transverse in average. Expressing the potential parts in Eq. (18) by the corresponding Fourier forms drives us closer to our goal, the Wigner-Liouville equation,

$$\partial_t W + p \cdot M^{-1} \cdot \partial_r W = \int_{-\infty}^{\infty} ik \cdot \partial_p <\delta(r-R)\delta(p-P)(\tilde{U} - \tilde{A} \cdot \dot{R})\exp(ik \cdot R)> d^{3N}(k/2\pi) \qquad (19)$$

Following the hint from Eq. (13), the random potential fluctuations $\tilde{A} \cdot \dot{R} = \tilde{\xi} \hbar k \cdot \dot{R}$ can be expressed by a global fluctuating parameter $\xi(R)$, being dimensionless and zero-centered. Substituting this expression in Eq. (19), the latter changes after simple rearrangements to

$$\partial_t W + p \cdot M^{-1} \cdot \partial_r W = \int_{-\infty}^{\infty} ik \cdot \partial_p <\delta(r-R)\delta(p-P)(\tilde{U} + \tilde{\xi} i\hbar \partial_t)\exp(ik \cdot R)> d^{3N}(k/2\pi) \qquad (20)$$

It is well known from electrodynamics that the gauge transformation $U \to U - \partial_t f$ and $A \to A + \partial_r f$ does not affect the physical state.[7] From $f = i\hbar \ln \Psi$, for instance, one can easily recognize the origin of the quantum mechanical operators of energy and momentum. The gauge theory explains how the force carriers transmit the potential interaction $U$ via the vector potential $A$. Appling the gauge transformation in Eq. (20) to forward the random potential fluctuations results in an alternative equation

$$\partial_t W + p \cdot M^{-1} \cdot \partial_r W = \int_{-\infty}^{\infty} ik \cdot \partial_p <\delta(r-R)\delta(p-M\cdot\dot{R}-A-\xi i\hbar\partial_R)\tilde{U}\exp(ik\cdot R)> d^{3N}(k/2\pi) \qquad (21)$$

and, using the properties of the Dirac delta function, Eq. (21) can be further elaborated to

$$\partial_t W + p \cdot M^{-1} \cdot \partial_r W = \int_{-\infty}^{\infty} \tilde{U}\partial_r \exp(ik\cdot r) \cdot \partial_p <\delta(r-R)\delta(p-P+\xi\hbar k)> d^{3N}(k/2\pi) \qquad (22)$$

In the deterministic case ($\xi=0$) Eq. (22) reduces to Eq. (14). Apparently, the symmetric fluctuations of $\xi(R)$ around zero generate the stochasticity of quantum mechanics. To reproduce Eq. (13), one should identify the probability density $W_k = <\delta(r-R)\delta(p-P+\xi\hbar k)>$, if the $\xi$-fluctuations are uniformly distributed in range $\pm 1/2$. Perhaps, the global parameter $\xi$ is related to the force carriers' helicity, which is an important quantum property of photon beams.[16,17] The present theory is a kind of stochastic electrodynamics (SED). The existing SED models describes, however, the fluctuations of the electromagnetic field in time, caused by the zero-point energy of vacuum.[18,19] The latter result always in energy changes and that is way they are excluded in the present model. To compensate the energy fluctuations, the classical SED models consider the dissipative Abraham-Lorentz force additionally in the stochastic Newton equation. How we showed already,[20] this results in a description of the quantum Brownian motion, not of quantum mechanics itself. Moreover, the Abraham-Lorentz force is relativistic, while the speed of light $c$ appears now for the first time in the present papers, since in the frames of a non-relativistic mechanics it should be set infinite. The zero-point energy fluctuations in vacuum are not accounted in the Schrödinger equation (1) and they cause in addition the Lamb shift[21] of the hydrogen orbitals. The size effects of the electron and proton are also neglected in Eq. (1) but they could result in momentum transfer via direct collisions of the point particles.[22]

## References

1. E. Schrödinger, An undulatory theory of the mechanics of atoms and molecules, *Phys. Rev.* **28** (1926) 1049-1070

2. R. Tsekov, Emerging quantum mechanics from stochastic dynamics of virtual particles, *J. Phys. Conf. Ser.* **701** (2016) 012034
3. E.P. Wigner, On the quantum correction for thermodynamic equilibrium, *Phys. Rev.* **40** (1932) 749-759
4. H. Weyl, The Theory of Groups and Quantum Mechanics, Dover, New York, 1950
5. H.J. Groenewold, On the principles of elementary quantum mechanics, *Physica* **12** (1946) 405-460
6. J.E. Moyal, Quantum mechanics as a statistical theory, *Proc. Cambridge Phil. Soc.* **45** (1949) 99-124
7. A. Zee, Quantum Field Theory in a Nutshell, Princeton University Press, Princeton, 2010
8. R. Tsekov, On the stochastic origin of quantum mechanics, *Rep. Adv. Phys. Sci.* **1** (2017) 1750008
9. R.L. Hudson, When is the Wigner quasi-probability density non-negative?, *Rep. Math. Phys.* **6** (1974) 249-252
10. F. Soto and P. Claverie, When is the Wigner function of multidimensional systems nonnegative?, *J. Math. Phys.* **24** (1983) 97-100
11. J. Weinbub and D.K. Ferry, Recent advances in Wigner function approaches, *Appl. Phys. Rev.* **5** (2018) 041104
12. W.B. Case, Wigner function and Weyl transforms for pedestrians, *Am. J. Phys.* **76** (2008) 937-946
13. S. Nouri, Wigner phase-space distribution function for the hydrogen atom, *Phys. Rev. A* **57** (1998) 1526-1528
14. P. Campos, M.G.R. Martins, M.C.B. Fernandes and J.D.M. Vianna, Quantum mechanics on phase space: The hydrogen atom and its Wigner functions, *Ann. Phys.* **390** (2018) 60-70
15. R. Tsekov, Brownian motion and quantum mechanics, *Fluct. Noise Lett.* **19** (2020) 2050017
16. K.Y. Bliokh and F. Nori, Transverse and longitudinal angular momenta of light, *Phys. Rep.* **592** (2015) 1-38
17. I. Bialynicki-Birula and Z. Bialynicka-Birula, Quantum-mechanical description of optical beams, *J. Opt.* **19** (2017) 125201
18. H.E. Puthoff, Ground state of hydrogen as a zero-point-fluctuation-determined state, *Phys. Rev. D* **35** (1987) 3266-3269
19. L. de la Pena and A.M. Cetto, The Quantum Dice: An Introduction to Stochastic Electrodynamics, Kluwer, Dordrecht, 1996
20. R. Tsekov, Brownian emitters, *Fluct. Noise Lett.* **15** (2016) 1650022
21. W.E. Lamb and R.C. Retherford, Fine structure of the hydrogen atom by a microwave method, *Phys. Rev.* **72** (1947) 241-243
22. R. Tsekov, Hard spheres model of the atom, *Chemistry* **24** (2015) 818-824